\documentclass[%
 preprint,
groupedaddress,
 amsmath,amssymb,
 aps,
floatfix,
]{revtex4-1}

\usepackage{graphicx}
\usepackage{dcolumn}
\usepackage{bm}
\usepackage{amsmath}
\usepackage{amssymb}
\usepackage{subfig}
\usepackage{float}

\usepackage{verbatim}
\usepackage{booktabs}
\usepackage{color}

\newcommand{\Michelle}[1]{{\color{black}#1}}
\newcommand{\modif}[1]{{\color{black}#1}}

\begin{document}


\title{A minimal model for a hydrodynamic fingering instability in microroller suspensions}%

\author{Blaise Delmotte}
\email{delmotte@courant.nyu.edu}
 \affiliation{Courant Institute of Mathematical Sciences,
New York University, New York, NY 10012, USA.}
\author{Michelle Driscoll}%
\affiliation{%
 Department of Physics, New York University, New York, NY 10003, USA.
}%

\author{Aleksandar Donev}
\affiliation{%
 Courant Institute of Mathematical Sciences,
New York University, New York, NY 10012, USA. 
}%

\author{Paul Chaikin}
\affiliation{
 Department of Physics, New York University, New York, NY 10003, USA.
}%

\begin{abstract}
We derive a minimal continuum model to investigate the hydrodynamic mechanism behind the fingering instability recently discovered  in a suspension of microrollers near a floor [Driscoll \textit{et al.} Nature Physics, 2016]. Our model, consisting of two  continuous lines of rotlets, exhibits a linear instability driven only by 
\Michelle{hydrodynamic} interactions,  and  reproduces the lengthscale selection observed in  large scale particle simulations and in  experiments.  By adjusting only one parameter, the distance between the two lines, our dispersion relation exhibits quantitative agreement with the simulations and qualitative agreement with experimental measurements. Our linear stability analysis indicates that this instability is caused by the combination of the  advective and  transverse flows generated by the microrollers near a no-slip surface. Our simple model offers an interesting formalism to characterize other hydrodynamic instabilities that have not been  yet well understood, such as size scale selection in suspensions of particles sedimenting adjacent to a wall, or the recently observed formations of traveling phonons in systems of confined driven particles.
\end{abstract}

\maketitle

 \section{Introduction}

When suspended in a viscous fluid at low Reynolds number, small moving particles interact through long range hydrodynamic  interactions. These many-body interactions can give rise to strong density and velocities fluctuations  in the bulk and lead to instabilities  \cite{Guazzelli2011}. The addition of boundaries strongly modify the hydrodynamic interactions between particles and affect the dynamics of the system. For instance, sedimenting particles between two parallel plates exhibit a transverse Rayleigh-Taylor-like instability whose wavelength strongly depends on the distance between the plates \cite{Carpen2002}.

When these particles are driven by means of an external field, or by self-propulsion mechanisms, they induce active flows that modify the interactions within the suspension and sometimes lead to strong  density fluctuations and long ranged orientational correlations \cite{Saintillan2013,Marchetti2013}. A well-known example is the instability of the isotropic state in suspensions of elongated swimmers, which was predicted by the theory \cite{AditiSimha2002, Saintillan2008a,Baskaran2009,Ezhilan2013} and reported  by both numerics \cite{Saintillan2007,Saintillan2011,Lushi2013} and experiments \cite{Wu2000,Sokolov2007,Wensink2012,ryan_2013,Creppy2015}. 
 Under confinement, active and driven  suspensions exhibit a wide variety of behaviours such as the formation of vortices, asters or polar bands \cite{Lefauve2014, Brotto2013, Tsang2015, Tsang2016}.

\begin{figure}
\includegraphics[width=1\textwidth]{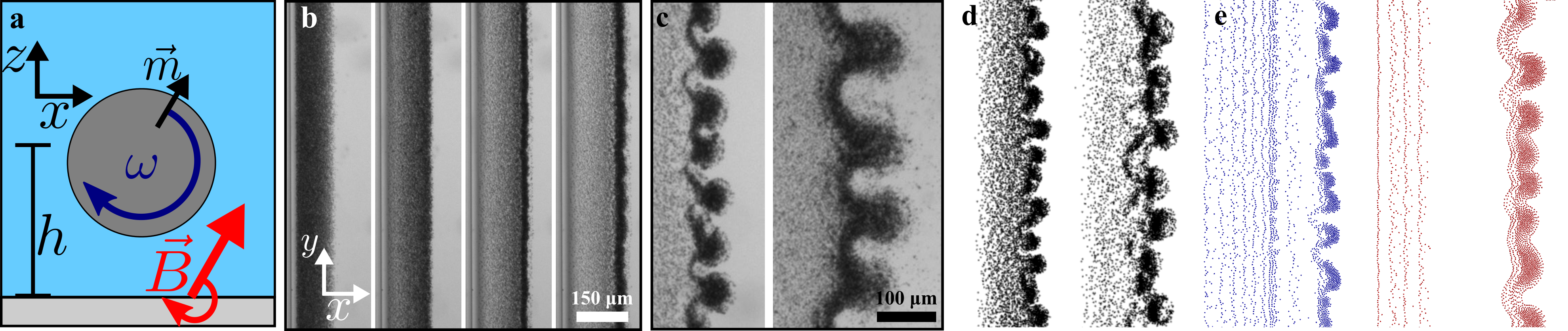}
\caption{\modif{\textbf{a} Schematic of a single microroller at height $h$ above the chamber floor.  Due to the permanent moment, $\mathbf{m}$, the particle can be rotated at angular frequency $\mathbf{\omega}$ by an external magnetic field $\mathbf{B}$ about the $y$-axis. \textbf{b} Experimental images of the shock front development from initial uniform state ($f = 20$ Hz, $h_g = 1\mu$m).}  \textbf{c} Experimental images of fingering instability at two different heights ($f = 10$ Hz) \cite{critters}. Left: $h_g = 1$ $\mu$m. Right: $h_g = 2.5$ $\mu$m.  \modif{\textbf{d} Large scale 3D Brownian dynamic simulations using same parameters as in the experiments shown on panel \textbf{c}  \cite{Usabiaga2017}}. \textbf{e} Quasi-2D  simulations without Brownian motion \cite{critters}, $f=6.4$ Hz. Left: particle height $ h= 1.97$ $\mu$m.  Right: $ h= 3.94$ $\mu$m. } 
\label{fig:fingers}
\end{figure}

In a recent work, we have uncovered a new hydrodynamic instability in a driven system of  microrollers: suspensions of colloids rotating  parallel to a floor \cite{critters}. 
\modif{These microrollers consists of \Michelle{spherical polymer} 
colloids with radius  $a=0.66$ $\mu$m which have a small permanent magnetic moment ($\lvert \mathbf{m}\rvert \sim 5\cdot10^{-16}$ Am$^2$) 
\Michelle{due to} an embedded hematite cube \cite{Sacanna2012}, see schematic in Fig.\ \ref{fig:fingers}a. Their equilibrium gravitational  height is  given by the competition between gravity and thermal fluctuations: $h_g = a + k_BT/mg$, where $a$ is the particle radius, $m$ its buoyant mass, $g$ the gravitational acceleration and $k_BT$ the thermal agitation energy.  More details about the experimental system are provided in \cite{critters}.}
When driven by a magnetic field rotating about an axis parallel to the floor, these microrollers generate strong advective flows in their vicinity. These flows are responsible for the large-scale collective effects observed in uniform suspensions. When the particle distribution is discontinuous, the microrollers form a shock front 
\modif{ with a well defined width (Fig.\ \ref{fig:fingers}b)}, which quickly becomes unstable and generates finger-like structures with compact tips \modif{(Fig.\ \ref{fig:fingers}c)}. These fingertips can detach to form compact autonomous  structures, called``critters" \cite{critters}.


Our large scale particle simulations and experiments suggest that the fingering instability is a linear instability controlled by the height of the particles above the floor \modif{(Fig.\ \ref{fig:fingers}c,d,e)}. 
\Michelle{With the addition of } Brownian motion, our simulations \modif{(Fig.\ \ref{fig:fingers}d)} achieve quantitative agreement with the experiments on the measured wavelength for various gravitational heights $h_g$ \cite{Usabiaga2017}. 
The fingering instability was also reproduced with a much simpler 2D system \modif{(Fig.\ \ref{fig:fingers}e)}: rotlets, i.e. point  torques, interacting only through hydrodynamic interactions in a plane at a fixed height above the floor \cite{critters}.


These simulations showed that the instability is controlled by far-field hydrodynamic interactions in the \textit{plane parallel to the floor}, but they  do not provide  a clear physical picture of the instability mechanism. 
Nonlinear phenomena such as the shock formation do not elucidate the exact mechanisms at play behind the transverse instability of the front that forms the fingers.

We recently developed a theoretical model to study density fluctuations in uniform suspensions of microrollers and to investigate the formation of the shock front \cite{Delmotte2017}.
This model relies on a nonlocal description of the \modif{far-field} hydrodynamic interactions between microrollers treated as rotlets. The theoretical results and comparisons with experiments and simulations showed \modif{that the front  forms due to the nonlocal nature of far-field hydrodynamic interactions in the plane above the floor, and that it has a well defined finite width which is controlled by the particle height.} However, the model used in this prior work to describe the front \cite{Delmotte2017} is one-dimensional and does not allow for transverse variations, and therefore cannot be used to study the two-dimensional transverse fingering instability.

In this paper, we derive a minimal two-dimensional continuum model, based on a nonlocal description, in order to:  (1) confirm that the transverse instability is a linear instability,  (2) study the dependence of the characteristic wavelength on the control parameters, (3)  identify precisely the  hydrodynamic mechanisms at play.  
We compare the linear stability analysis with numerical simulations and experiments and discuss the validity of the model. We take advantage of the flexibility and simplicity of the model to study the stability of the microrollers above a slip surface and finally offer promising extensions to other particulate systems within the same framework.

 


 \section{Model and linear stability analysis}


Our goal is to model the shock front in a simplified manner in order to carry out an analytic linear stability analysis of the system. Using a finite sheet would make the calculations untractable because of the nonlocality of the hydrodynamic interactions. 
Suspensions of particles have been studied with a two-fluid  approach \cite{Carpen2002,Pan2001, Wysocki2009}, which requires computing the effective viscosity of the  suspension. However, a two-fluid model is not applicable to the confined microroller suspension where hydrodynamic interactions are nonlocal and depend on the particle height $h$. 
We found that the simplest and most relevant approach to analytically model an unstable front of microrollers is to consider \Michelle{\emph{two lines}} of rotlets interacting hydrodynamically in a plane parallel to the floor at a fixed height $h$, \Michelle{as one line of rotlets is linearly stable to perturbations.}


\subsection{Governing equations}
\begin{figure}
\subfloat[][]{\includegraphics[width=0.64\textwidth]{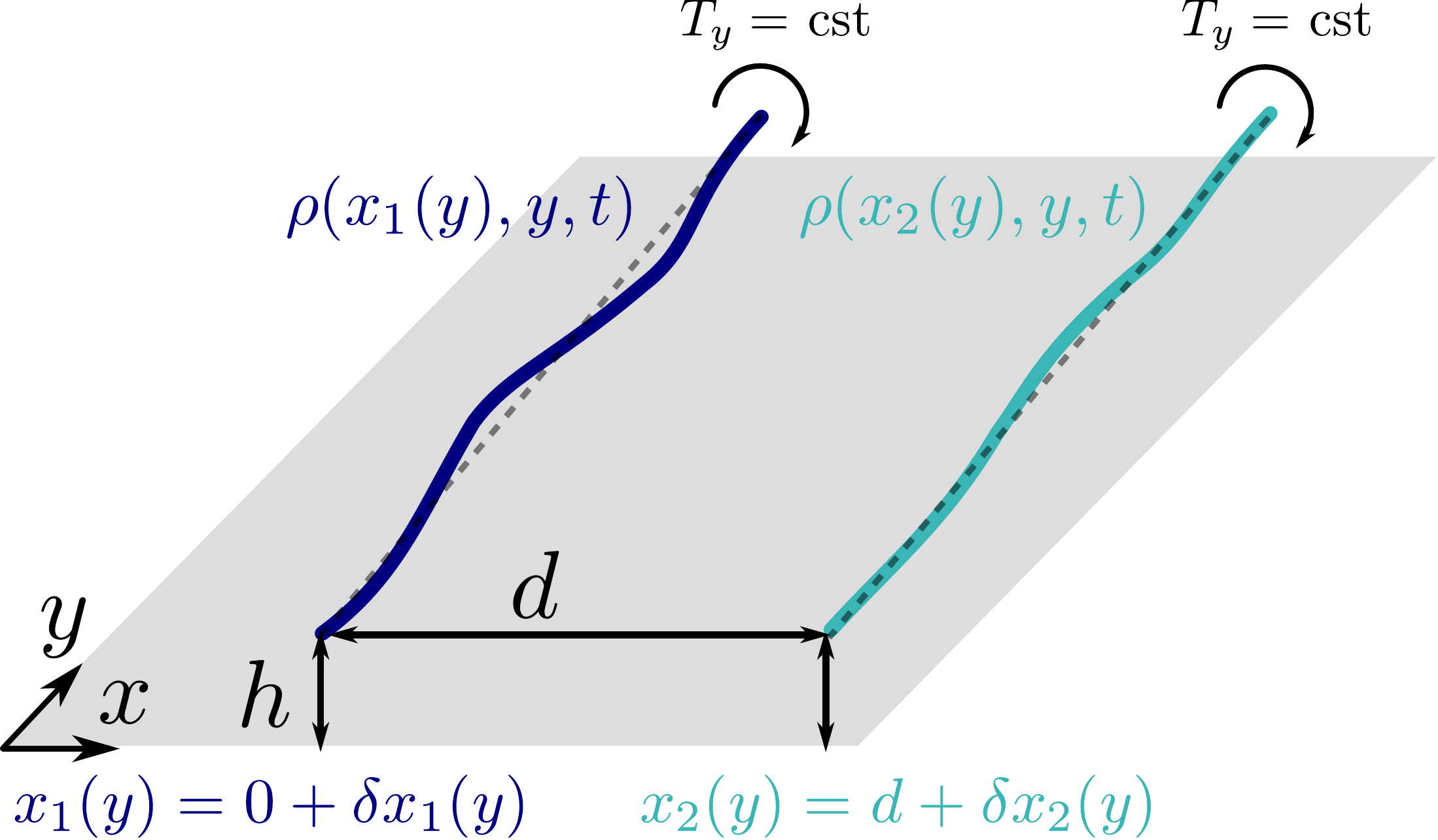}}
\subfloat[][]{\includegraphics[width=0.32\textwidth]{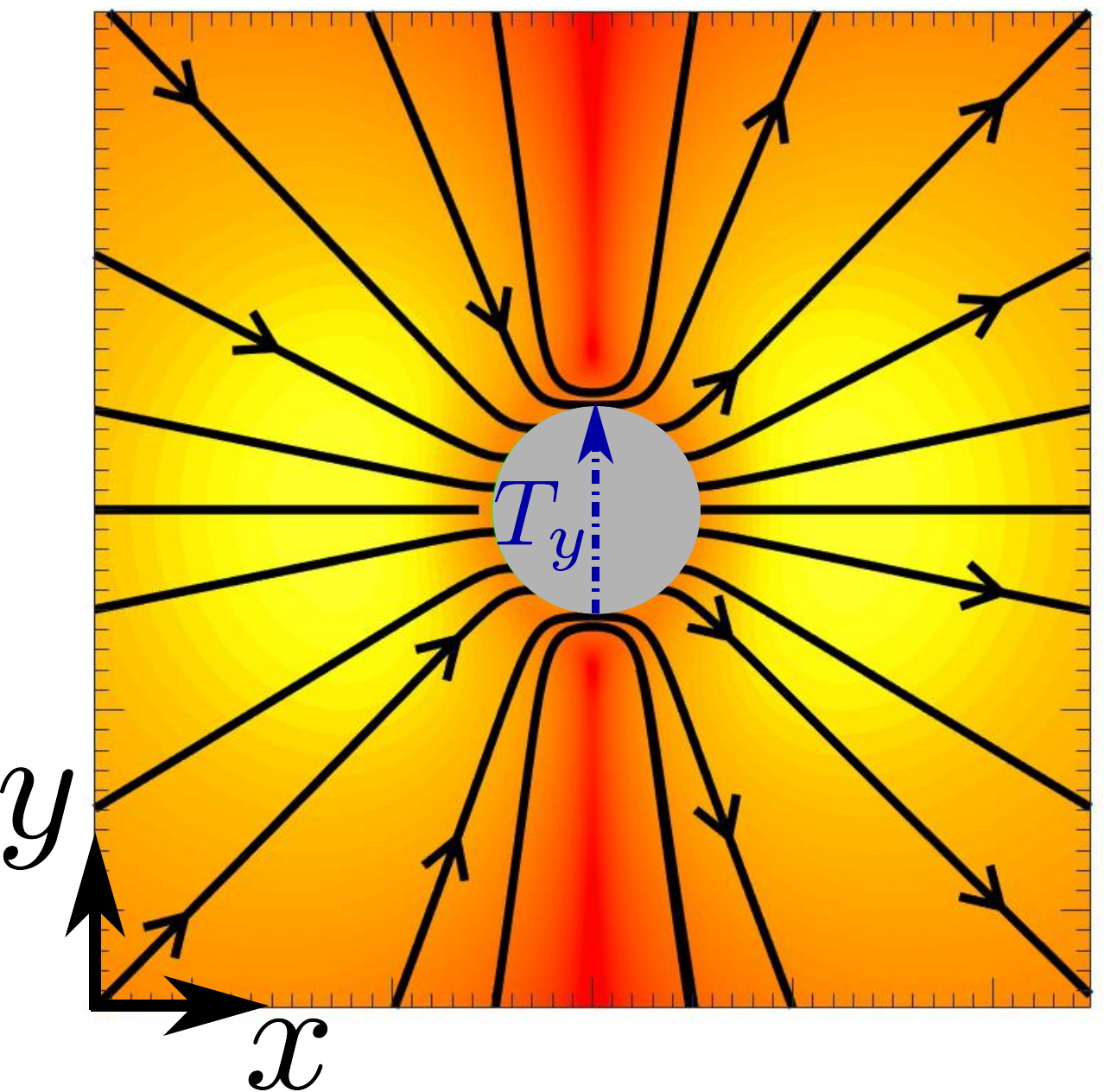}}
\caption{(a) Sketch of the two infinite lines. The rotlet densities are rotating with a constant  
torque, $T_y$, about the $y$-axis and interact hydrodynamically in the $xy$-plane at a height $h$ above the floor. (b) Flow field around a rotating sphere in the $xy$-plane of rotation at a distance $h$ above a no-slip surface \cite{critters}. The arrow indicates the axis of the applied torque $T_y$. The colors, from red to yellow, represent the magnitude of the in-plane velocity.} 
\label{fig:sketch_lines}
\end{figure}

Consider two infinite lines of rotlets, labelled ``1'', at the back, and ``2'', at the front, respectively, with rotlet densities $\rho_1(y)$ and $\rho_2(y)$,  rotating around the $y$ axis in a plane at a fixed height $h$ above a no-slip boundary. Their $x$ position is parametrized by $y$, with an initial state  $x_1(y,t=0) = 0$ and $x_2(y,t=0) =d$  (see Fig.\ \ref{fig:sketch_lines}a).
Here, we neglect out of plane motion in the $z$-direction and only consider the velocities in the $xy$-plane.  Although out of plane motion is seen in the experiments, our quasi-2D simulations in Fig.\ \ref{fig:fingers}c confirm that the fingering instability still occurs when considering  particles confined to a plane (with 3D hydrodynamics).

A point rotlet located at a position $(x',y';\,h)$ induces a \modif{horizontal} fluid velocity at point $(x,y;\,h)$  given by \cite{Blake1974, Delmotte2017}:
\begin{eqnarray}
v_{x}(x,y;\,h) &=& \mathcal{G}_x\left(x-x',y-y';\,h\right) \nonumber\\
&=&  Sh\tfrac{(x-x')^2}{\left[(x-x')^2 + (y-y')^2 + 4h^2\right]^{5/2}}, \label{eq:int_kernelx}
\end{eqnarray}
\begin{eqnarray}
v_{y}(x,y;\,h) &=& \mathcal{G}_y\left(x-x',y-y';\,h\right) \nonumber\\
&= & Sh\tfrac{(x-x')(y-y')}{\left[(x-x')^2 + (y-y')^2 + 4h^2\right]^{5/2}}, \label{eq:int_kernely}
\end{eqnarray}
where $S = 6T_y/(8\pi\eta)$, $T_y>0$  is the applied constant torque around the $y$-axis (i.e. the rotlet magnitude). Changing the sign of $T_y$ would change the direction of motion of the system and reverse the role of the lines. $\mathcal{G}_x$ and $\mathcal{G}_y$ are the hydrodynamic interaction kernels in the $xy$-plane. Note that these kernels are not singular because of the finite constant height in the denominator $h>0$. The  velocity field around a rotating particle above a no-slip wall is shown in Fig.\ \ref{fig:sketch_lines}b. In an unbounded fluid, this velocity field would be zero in the whole plane. The transverse flow $v_y$ observed in Fig.\ \ref{fig:sketch_lines}
\Michelle{b} arises from the image system of the rotlet which includes a stresslet \cite{Blake1974}. 
Since $h$ is fixed, we will henceforth omit it in the argument of the kernels:  $\mathcal{G}(x,y) \equiv   \mathcal{G}(x,y;h)$.

The velocity induced by  the front line ``2" at a point $(x_1(y),y)$  on the back line ``1" is given by the functionals
\begin{eqnarray}
  V^{(1,I)}_{x}\left(x_1(y),y\right)  &=&  \int\limits_{-\infty}^{+\infty}   \mathcal{G}_x\left(x_1(y)-x_2(y'),y-y'\right) \rho_2(y')dy' \\
 V^{(1,I)}_{y}\left(x_1(y),y\right)  &=&  \int\limits_{-\infty}^{+\infty}   \mathcal{G}_y\left(x_1(y)-x_2(y'),y-y'\right) \rho_2(y')dy',   
\end{eqnarray}
where $\rho_2(y')=\rho(x_2(y'),y')$.
In $V^{(1,I)}_{x,y}$ the first superscript ``1'' indicates the line considered and the second superscript ``I'' stands for ``induced'' by the other line.
The self-induced velocity, ``S'', is given by
\begin{eqnarray}
 V^{(1,S)}_{x}\left(x_1(y),y\right) &=& \int\limits_{-\infty}^{+\infty}   \mathcal{G}_x\left(x_1(y)-x_1(y'),y-y'\right) \rho_1(y')dy'  \\
V^{(1,S)}_{y}\left(x_1(y),y\right) &=& \int\limits_{-\infty}^{+\infty}   \mathcal{G}_y\left(x_1(y)-x_1(y'),y-y'\right) \rho_1(y')dy'  .    
\end{eqnarray}
Similar terms are derived for line 2. 
The governing equations for the two lines, in the limit of small deviations from a straight line, are the equations of motion in the direction perpendicular to the lines 
\begin{eqnarray}
\dfrac{\partial x_1(y,t)}{\partial t} =  V^{(1,S)}_{x} + V^{(1,I)}_{x}, \label{eq:motion_1}\\
\dfrac{\partial x_2(y,t)}{\partial t} =  V^{(2,S)}_{x} + V^{(2,I)}_{x},
\end{eqnarray}
and  mass conservation coming from the motion parallel to the lines,
\begin{eqnarray}
\dfrac{\partial \rho_1(y,t)}{\partial t} = -\dfrac{\partial \left[\rho_1(y,t)\left( V^{(1,S)}_{y} + V^{(1,I)}_{y} \right)\right]}{\partial y},\\
\dfrac{\partial \rho_2(y,t)}{\partial t} = -\dfrac{\partial \left[\rho_2(y,t)\left( V^{(2,S)}_{y} + V^{(2,I)}_{y} \right)\right]}{\partial y}. \label{eq:continuity_2}
\end{eqnarray}

These four nonlinear, nonlocal, coupled equations \eqref{eq:motion_1}-\eqref{eq:continuity_2} can be linearized about the homogeneous state.

\subsection{Linear stability analysis}
We perturb the $x$ positions of each line about their initial position 
\begin{eqnarray}
 x_1(y,t)  =  0 + \delta x_1(y,t)\\ 
 x_2(y,t)  =  d + \delta x_2(y,t),
 \end{eqnarray}
 where $\delta x_1, \delta x_2\ll \min(d,h)$, and their rotlet densities about a constant value $\rho_0$, 
 \begin{eqnarray}
 \rho_1(y,t)  =  \rho_{0} + \delta \rho_1(y,t)\\
 \rho_2(y,t)  =  \rho_{0} +  \delta \rho_2(y,t),
\end{eqnarray}
where $\delta \rho_1, \delta \rho_2 \ll \rho_0$.
After Taylor expanding the functionals in Eqs.\ \eqref{eq:motion_1}-\eqref{eq:continuity_2} we obtain the linearized governing equations
\begin{eqnarray}
\dfrac{\partial \delta x_1}{\partial t} &=& \rho_0 \mathcal{G}_0\delta x_1 +  \rho_0 \mathcal{G}_1*\delta x_2  +  \mathcal{G}_{x}(0-d,\cdot)*\delta\rho_2 \label{eq:adv_linearized}\\
\dfrac{\partial \delta x_2}{\partial t} &=& -\rho_0 \mathcal{G}_0\delta x_2 - \rho_0 \mathcal{G}_1*\delta x_1  +  \mathcal{G}_{x}(0-d,\cdot)*\delta\rho_1\label{eq:adv_linearized2}\\
\dfrac{\partial \delta\rho_1}{\partial t} &=& -\rho_0\dfrac{\partial \left[ \left(\rho_0 \mathcal{G}_2*\delta x_1 + \rho_0 \mathcal{G}_3*\delta x_2  +  \mathcal{G}_{y}(0-d,\cdot)*\delta\rho_2 \right)\right]}{\partial y}\\
\dfrac{\partial \delta\rho_2}{\partial t} &=& -\rho_0\dfrac{\partial \left[ \left(\rho_0 \mathcal{G}_2*\delta x_2 + \rho_0 \mathcal{G}_3*\delta x_1  -  \mathcal{G}_{y}(0-d,\cdot)*\delta\rho_1 \right)\right]}{\partial y},\label{eq:continuity2_linearized}
\end{eqnarray}
where we have used the symmetries of $\mathcal{G}_x$ and $\mathcal{G}_y$. The star ``$*$" denotes the one dimensional convolution product. 
The interaction kernels are given by 
\begin{eqnarray}
\mathcal{G}_0 &=& \left.\dfrac{\partial}{\partial x_1} \left[\int\limits_{-\infty}^{+\infty} \mathcal{G}_{x}\left(x_1-d,y-y'\right)  dy'\right]\right\rvert_{x_1=0} =   Sh \dfrac{8d(d^2-4h^2)}{3[d^2 + 4h^2]^{3}}, \nonumber\\
\mathcal{G}_1(y-y') &=& \left.\dfrac{\partial \mathcal{G}_x(0-x_2,y-y')}{\partial x_2}\right\rvert_{x_2 = d} =  -Shd\dfrac{3d^2-2[(y-y')^2+4h^2]}{\left[d^2+ (y-y')^2 + 4h^2\right]^{7/2}}, \nonumber\\
\mathcal{G}_2(y-y') &=& \left.\dfrac{\partial \mathcal{G}_y(0-x_1,y-y')}{\partial x_1}\right\rvert_{x_1 = 0} = -Sh(y-y')\dfrac{(y-y')^2+4h^2}{\left[ (y-y')^2 + 4h^2\right]^{7/2}}, \label{eq:G2}\nonumber\\
\mathcal{G}_3(y-y') &=& \left.\dfrac{\partial \mathcal{G}_y(0-x_2,y-y')}{\partial x_2}\right\rvert_{x_2 = d} =  Sh(y-y')\dfrac{4d^2-(y-y')^2-4h^2}{\left[d^2+ (y-y')^2 + 4h^2\right]^{7/2}}.\nonumber
\end{eqnarray}
Note that $\mathcal{G}_0$ is a constant which depends only on the geometric parameters of the problem $d$, $h$, and on $S$. Its physical meaning will be explained below. 

Looking for  periodic solutions of Eqs.\ \eqref{eq:adv_linearized}-\eqref{eq:continuity2_linearized} of the form
$$\mathbf{u} = (\delta x_1, \delta x_2, \delta \rho_1, \delta\rho_2) = \tilde{\mathbf{u}} \; e^{iky+\sigma t},$$
where the wavenumber $k = 2\pi/\lambda$ and
$$\tilde{\mathbf{u}} = (\delta \tilde{x}_1, \delta \tilde{x}_2, \delta \tilde{\rho}_1, \delta \tilde{\rho}_2),$$
we obtain the following eigenvalue problem
\begin{equation}
  A\tilde{\mathbf{u}} = \sigma \tilde{\mathbf{u}}
  \label{eq:EV_problem}
\end{equation}
where
\begin{equation}
   A=  \left[\begin{array}{cccc}
\rho_0\mathcal{G}_0 & \rho_{0}\tilde{\mathcal{G}}_1 & 0 & \tilde{\mathcal{G}}_{x}(-d,k)\\
-\rho_{0}\tilde{\mathcal{G}}_1 & -\rho_0\mathcal{G}_0 & \tilde{\mathcal{G}}_{x}(-d,k) & 0\\
-ik\rho_{0}^{2}\tilde{\mathcal{G}}_2 & -ik\rho_{0}^{2}\tilde{\mathcal{G}}_3 & 0 & -ik\rho_{0}\tilde{\mathcal{G}}_{y}(-d,k)\\
-ik\rho_{0}^{2}\tilde{\mathcal{G}}_3 & -ik\rho_{0}^{2}\tilde{\mathcal{G}}_2 & ik\rho_{0}\tilde{\mathcal{G}}_{y}(-d,k) & 0
\end{array}\right]. \label{eq:MatA}
\end{equation}

The complete expression for the entries of $A$ can be found in  Appendix A. 
Note that a model with one line of rotlets would be linearly stable since all self-induced terms are at least quadratic \modif{in the magnitude of the perturbations}. 

\subsubsection{Structure of the instability}

\begin{figure}
\centering
\subfloat[][Back line (line 1), located at  $x_1=0$.]{\includegraphics[width=0.49\textwidth]{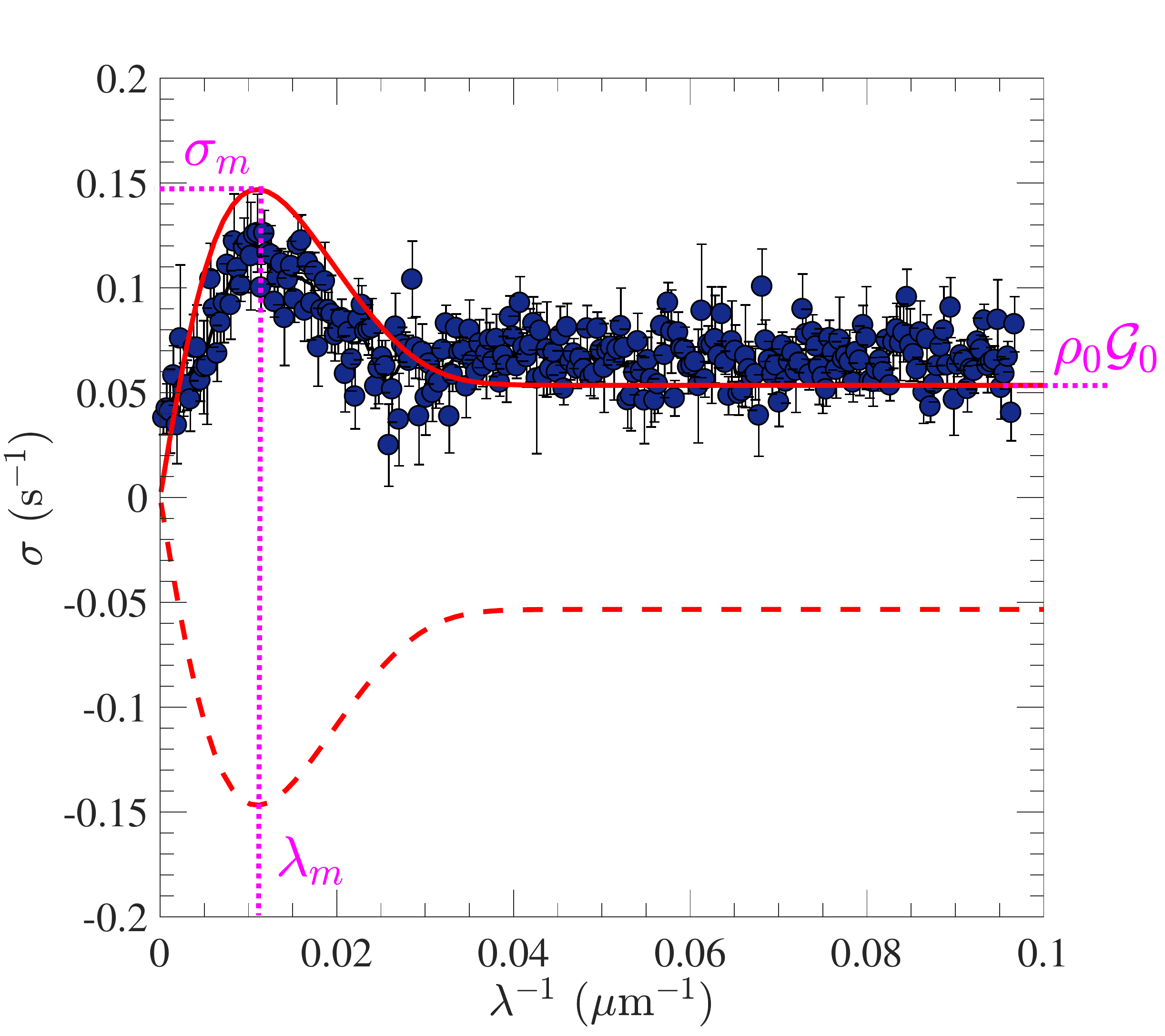}}
\hspace{0.2cm}
\subfloat[][Front line (line 2), located at $x_2=d$.]{\includegraphics[width=0.49\textwidth]{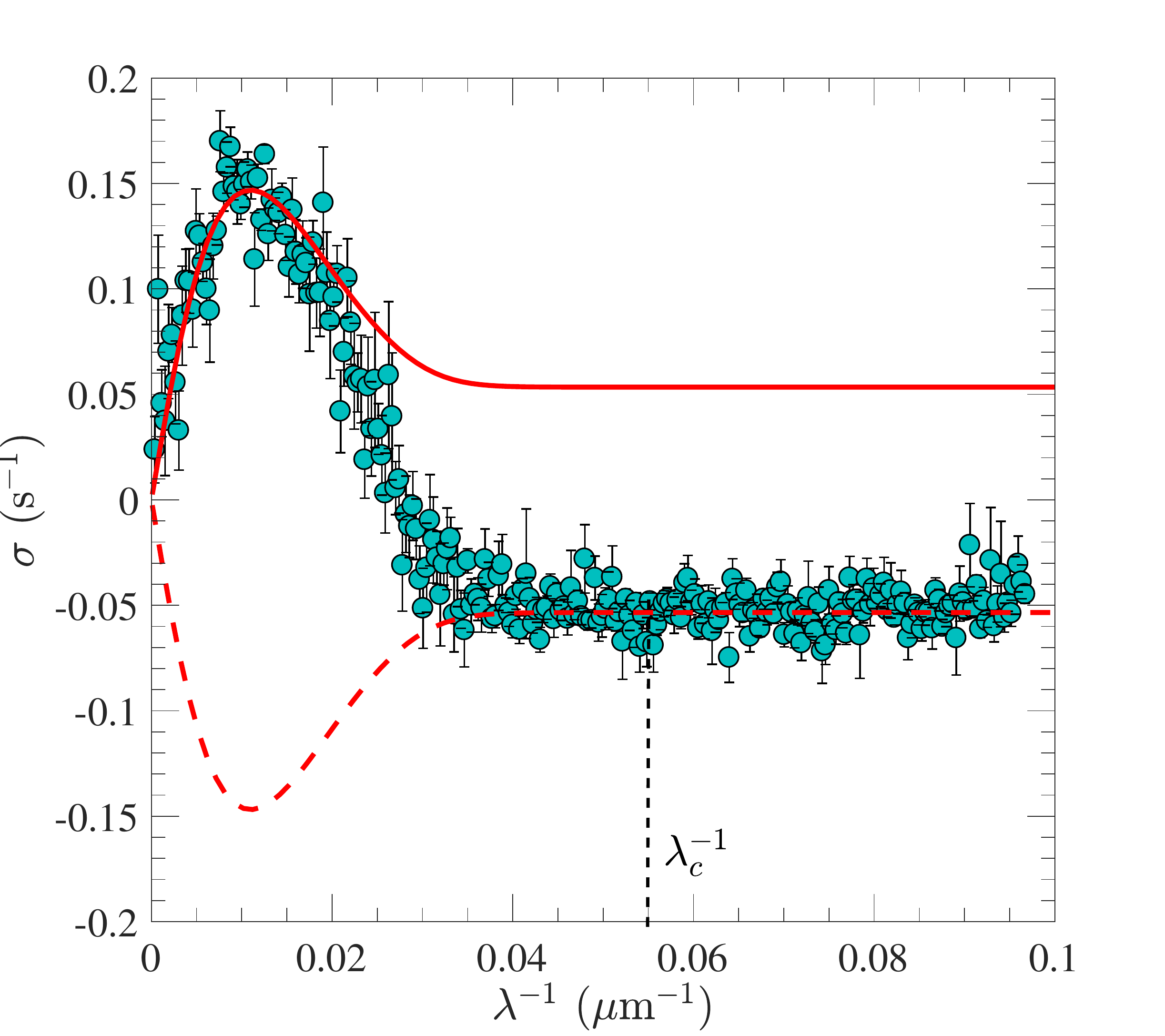}}\\
\subfloat[][]{\includegraphics[width=0.50\textwidth]{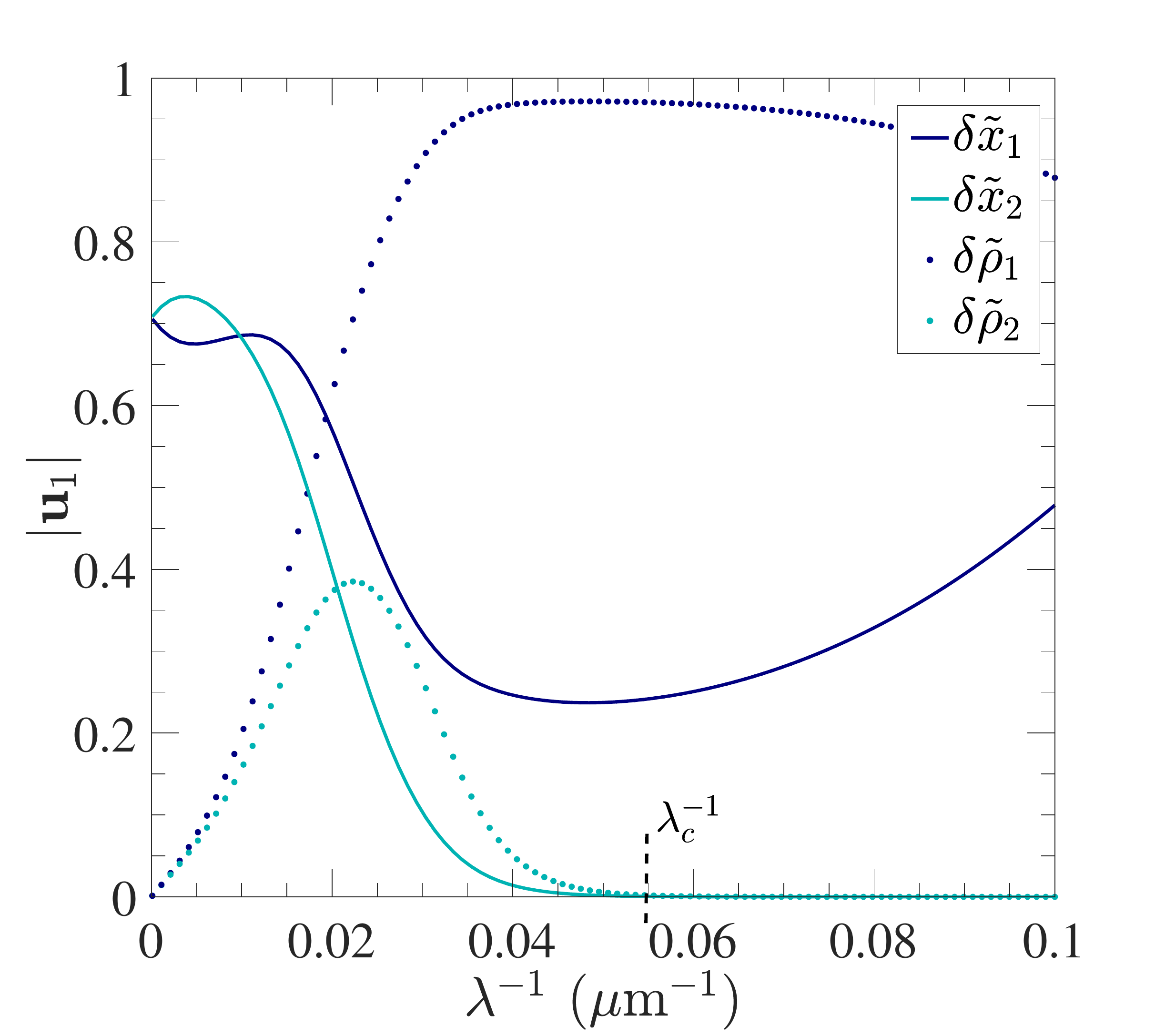}}
\hspace{0.2cm}
\subfloat[][]{\includegraphics[width=0.43\textwidth]{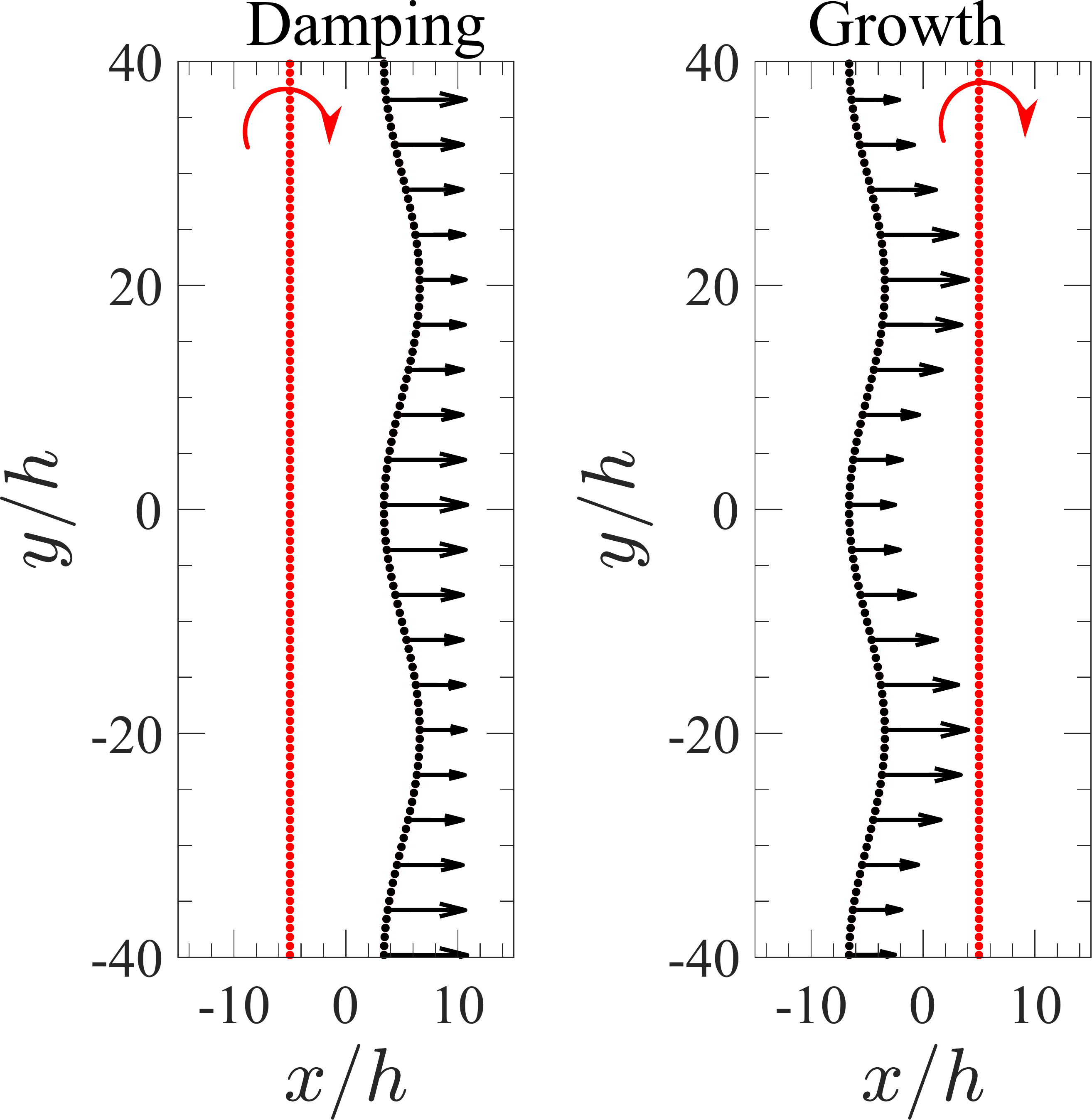}}
\caption{(a)-(b): Growth rate vs. $\lambda^{-1}$ for $h = 3.94$ $\mu$m and $d=10h$. Red lines: real eigenvalues $\sigma_{1}$ (solid) and $\sigma_2$ (dashed) from the two-line model (Eq. \eqref{eq:EV_problem}). 
Filled circles: growth rate obtained from simulation results averaged over 10 realizations. (c) Modulus of the entries in the amplified eigenmode $\mathbf{u}_1$ corresponding to the real positive eigenvalue $\sigma_1$ of the two-line model. (d) Velocity induced by a line of rotlets (red) on a perturbed passive line (black).  The red curved arrow indicates the direction of rotation. Left:  active line damps all modes of the passive line in front of it at a rate $-\rho_0\mathcal{G}_0$. Right: active line amplifies all modes of the passive line in the back of it at a rate $\rho_0\mathcal{G}_0$.} 
\label{fig:eigen_line}
\end{figure}

Due to the particular structure of $A$, Eq.\ \eqref{eq:EV_problem}  can be solved analytically. The four solutions $\sigma_1(k),..,\sigma_4(k)$ are written in Appendix A. The first two eigenvalues are real and of opposite sign $\sigma_1 = -\sigma_2$. The two other are imaginary and conjugate  $\sigma_3 = \bar{\sigma_4}$. All eigenvalues depend linearly on $S\propto T_y$. 
Figure \ref{fig:eigen_line}a shows the real eigenvalues $\sigma_{1,2}$ for $d = 10h$. First, one can see that the eigenvalues exhibit a well defined peak at $\lambda = \lambda_m$, meaning that the two-line model selects a fastest growing mode, which is characteristic of a  fingering instability.  Second, all modes, except the zero mode, are unstable and a clear plateau is visible for short wavelengths. The value of this plateau is exactly $\rho_0 \mathcal{G}_0$, which is the only constant entry in the matrix $A$. To better understand the presence of the plateau, we study the structure of the amplified eigenmode $\mathbf{u}_1$, associated to the real positive eigenvalue $\sigma_1$,  in Fig.\ \ref{fig:eigen_line}c. For this eigenmode, the density and position disturbances of the front line, $\delta \tilde{\rho}_2$ and $\delta \tilde{x}_2$, vanish for  $\lambda \approx \lambda_c$, where $\lambda_c$ is a critical value, while the back line disturbances, $\delta \tilde{\rho}_1$ and $\delta \tilde{x}_1$, are nonzero for all $\lambda$'s.

 This difference in behavior between the two lines can be understood by looking at the flow induced by one line on the other. Fig.\ \ref{fig:eigen_line}d   shows that a straight line of rotlets at $x=0$ \Michelle{(the back line)}, with uniform density $\rho_0$, damps the disturbances of the other line at $x=d$, regardless of their wavelength $\lambda$. 
 \Michelle{Conversely}, a straight line at $x=d$ \Michelle{(the front line)} increases the disturbances of the line at $x=0$ for all $\lambda$. 
 \modif{This phenomenon can be explained with a simple analogy: consider a force that dies off quickly with distance from a line. Take a parallel line a distance away and introduce a distortion so that parts are closer and others further away from the driving force. If the forces are attractive then the closer parts are drawn faster than the further parts and the distortion is amplified. If the forces are repulsive the closer parts are pushed away faster than the far ones and the distortion is reduced. For two lines of rollers the leading line advects the trailing one toward it while the tailing line pushes the leader away (Fig.\ \ref{fig:eigen_line}d). \Michelle{Thus,} the front line is stable while the rear goes unstable.}
 The term which relates a straight line with density $\rho_0$ and a line with perturbed position is precisely $\rho_0\mathcal{G}_0$ (see Eqs.\ \eqref{eq:adv_linearized}-\eqref{eq:adv_linearized2}). Thus the presence of the plateau is due to the growth of each mode on the back line. The term $\rho_0\mathcal{G}_0$  is compensated by $\rho_0\tilde{\mathcal{G}}_1(k)$ when $k\rightarrow 0$. Indeed $\lim_{k\rightarrow 0}  \rho_0\tilde{\mathcal{G}}_1(k) = - \rho_0Sh\frac{8d(d^2-4h^2)}{3(d^2+4h^2)^3} = -\rho_0\mathcal{G}_0$.

\subsubsection{Comparison with nonlinear simulations}

\begin{figure}
\centering
\includegraphics[width=0.99\textwidth]{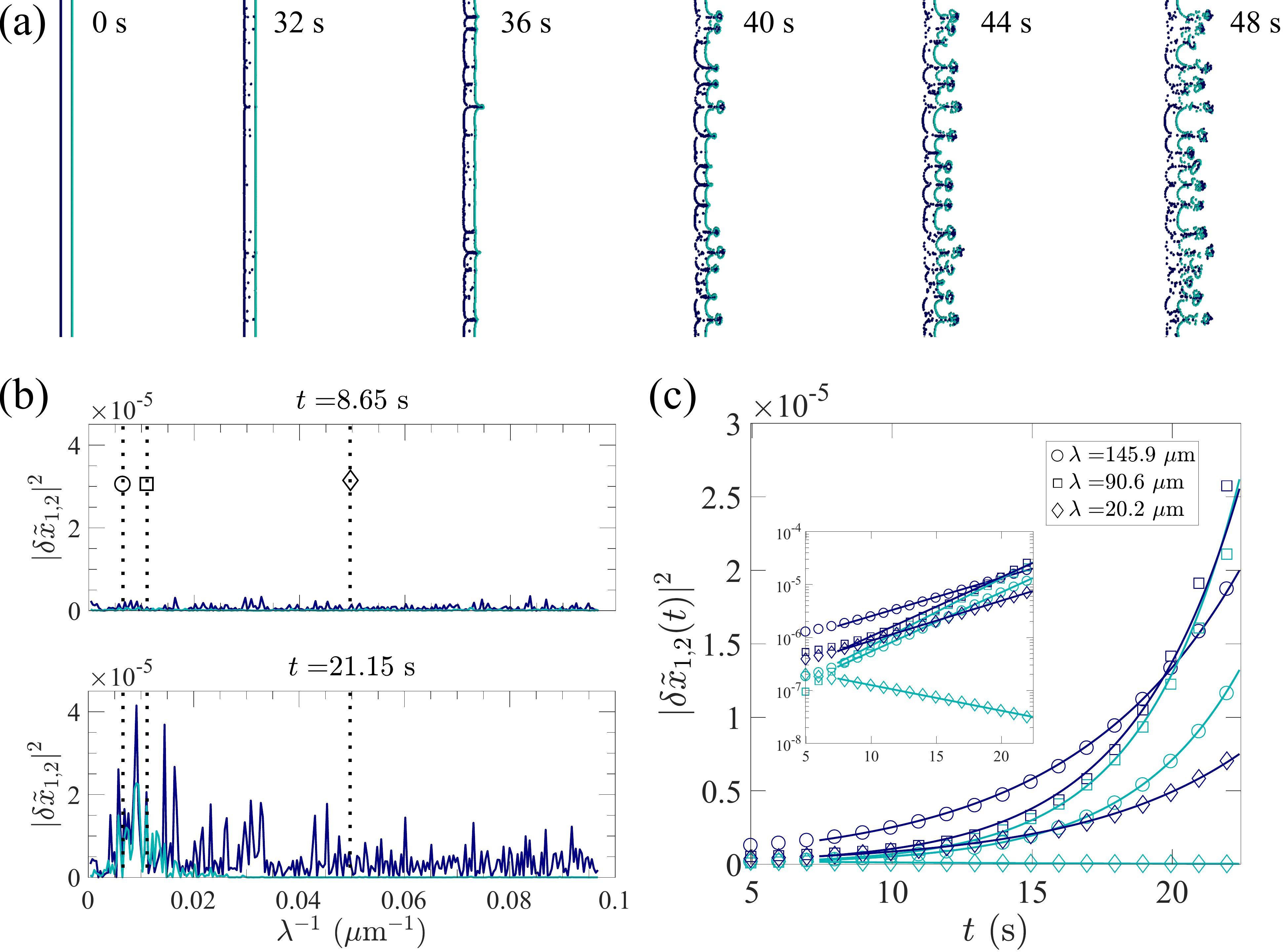}
\caption{Development of the instability in the nonlinear simulations with discrete rotlets for $h = 3.94$ $\mu$m and  $d = 10h$. Dark blue: back line. Light blue: front line (a) The time evolution of the two lines in real space at times $t=0 - 48$ s. (b) Fourier spectrum of the $x$ disturbances for each line at $t = 8.65$ s (top) and $t = 21.15$ s (bottom). The dotted vertical lines indicate the modes plotted in (c). (c)  Time evolution of the power of three Fourier modes $\lambda = 20, 91, 146$ $\mu$m for each line. Symbols: simulation data. Lines: exponential fits starting at $t = 8$ s. Inset: same data with a logarithmic scale on the $y$ axis.  }
\label{fig:rotlet_instability}
\end{figure}

We simulate the two-line system with discrete rotlet simulations.  Each line is discretized with 2000 rotlets interacting in the $xy$ plane.  The two  initially straight lines are perturbed by adding a small random increment to the particle positions.  The hydroynamic  interactions are calculated using Eqs.\ \eqref{eq:int_kernelx}-\eqref{eq:int_kernely}, and pseudo-periodic boundary conditions are used \cite{Usabiaga2017}. Positions are updated with an Adams-Bashforth-Moulton predictor-corrector scheme (see Methods section in \cite{critters}).
The full development of the fingering instability in the simulations is shown in Fig.\ \ref{fig:rotlet_instability}a.
To characterize the instability, we measure the position disturbance of each line $\delta x_{1,2}(y,t)$ that we bin along the $y$ direction with 1024 points. The resulting two  vectors are Fourier transformed to obtain the power spectra for each line shown in Fig.\ \ref{fig:rotlet_instability}b at two different times. At $t = 8.65$ s, both lines have a uniformly distributed spectrum. At $t = 21.15$ s, both spectra exhibit a clear peak at long wavelengths. The front line (light blue) has damped all the modes with short wavelength while the back line (dark blue) has amplified all modes. The time evolution of three Fourier modes $\lambda = 20, 91, 146$ $\mu$m is shown in Fig.\ \ref{fig:rotlet_instability}c. One can see that the shortest wavelength, $\lambda = 20$ $\mu$m, decreases for the front line and grows for the back line. We use an exponential fit $\delta \tilde{x}(t)\sim\exp(\sigma t)$ (see inset) to extract the growth rate $\sigma$ of each mode  for each line at the onset of the instability.

Fig.\ \ref{fig:eigen_line}a-b compare\Michelle{s} the extracted growth rate,  $\sigma(k)$, of each line, averaged over 10 simulations, to the theoretical predictions, with no 
\Michelle{adjustable} parameter\Michelle{s}.  
The theory is in excellent agreement with the particle simulations, 
\Michelle{demonstrating} that the linear stability analysis correctly captures the early-time dynamics of the fingering instability. 
Consistently with the theoretical predictions in Fig.\ \ref{fig:eigen_line}c, all modes of the back line are unstable ($\sigma>0$), while short wavelength disturbances on the front line are damped at a rate $-\rho_0\mathcal{G}_0$ when $\lambda$ is  below the critical value $\lambda_c$.


\subsubsection{Dependence on the control parameters}

\begin{figure}
\centering
\includegraphics[width=0.6\textwidth]{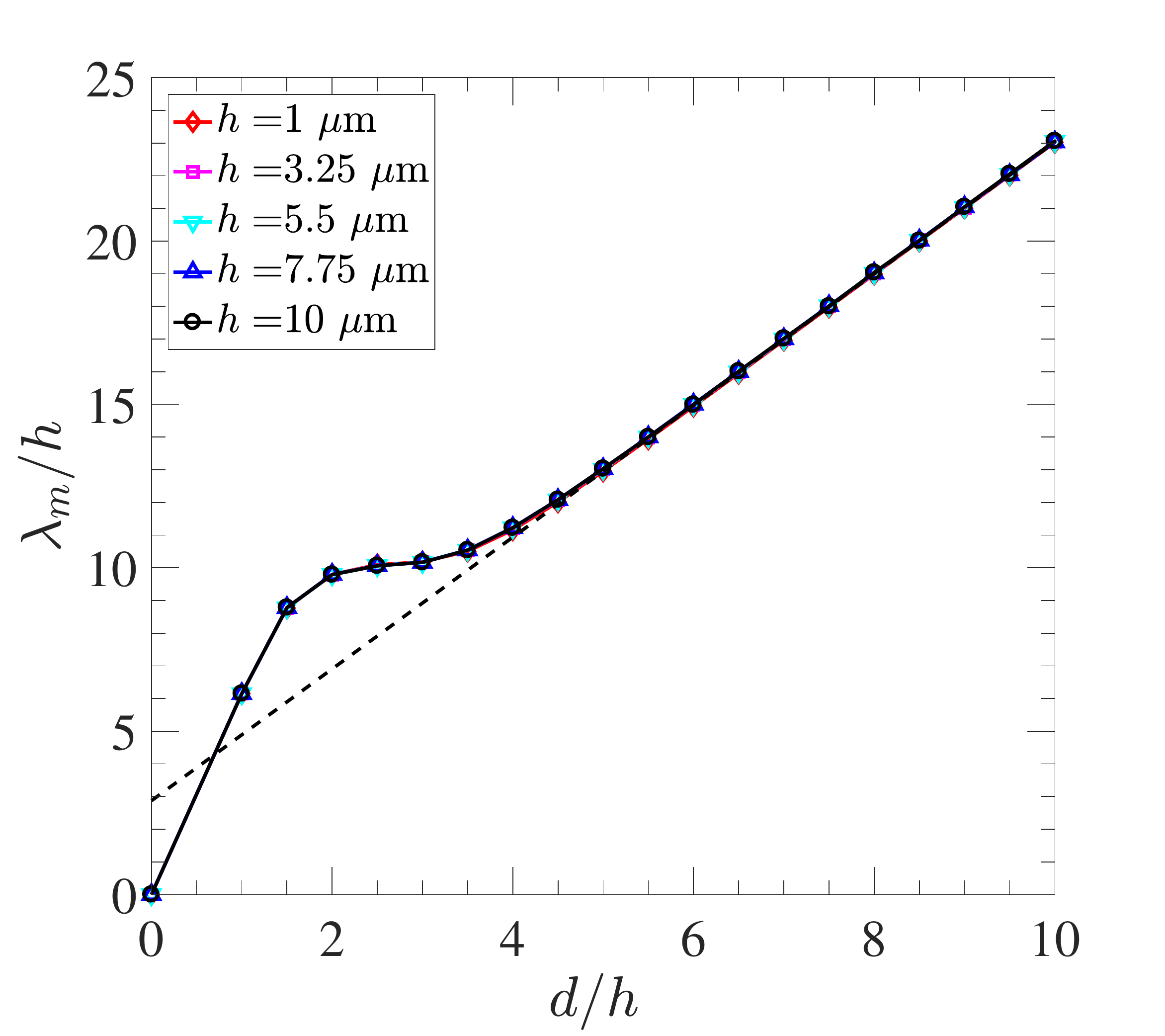}
\caption{$\lambda_m/h = f(d/h)$ vs. $d/h$ for various heights $h = 1 - 10$ $\mu$m. Solid line: numerical solution. Dashed line: linear fit $f(d/h) \approx 2.02(d/h)+2.85$ for $d/h>5$.} 
\label{fig:lambda_vs_h_vs_d}
\end{figure}

The fastest growing mode of the two line model  $\lambda_{m}$  is controlled by two geometric parameters: the height $h$ and the distance between the two lines $d$. 
Since $\sigma_1 = S f(h,d,k=2\pi/\lambda)$ has a complex analytic form (see Eq.\ \eqref{eq:EVS}), it is not possible to express $\lambda_m$ as a function of $h$ and $d$ explicitly. Instead we use numerical evaluations to determine $\lambda_m$ as a function of $h$ and $d$. Figure   \ref{fig:lambda_vs_h_vs_d} shows the graph of $\lambda_m/h$ vs. $d/h$ for various heights $h = 1-10$ $\mu$m. The collapse of the curves show that $\lambda_m$ depends linearly on the height $h$:  $\lambda_m  =   h f(d/h)$, where  $f(d/h)$ is a function of $d/h$. As shown in this Figure, $f(d/h)$, becomes approximately linear when $d/h>5$: $f(d/h) \approx 2.02(d/h) + 2.85$.

\section{Comparison with large scale simulations and experiments}

In this section we compare the two-line continuum model with numerical simulations and  experimental measurements.
We want to evaluate how relevant our model is to the more realistic microroller system.

The quasi-2D large scale simulations are performed with the method described in \cite{critters}. 
From the particle positions at a given time $t$ we can compute the empirical number density $n(x, y, t)$ in the $xy$-plane.  We compute the Fourier transform of $n(x, y, t)$
at $k_x = 0$, i.e., the Fourier transform $\tilde{n}(k_y = 2\pi/\lambda, t)$ of the number density along the direction of the front. We only use particles in the shock front, specifically, we only include the 70$\%$ of the particles with the largest $x$-coordinates. This ensures that the Fourier modes are not affected by the particles left behind the shock front. 
 We have confirmed that essentially the same results are obtained when including between 50$\%$ and 90$\%$ of the particles. 
 
\begin{figure}
\subfloat[][]{\includegraphics[width=0.47\textwidth]{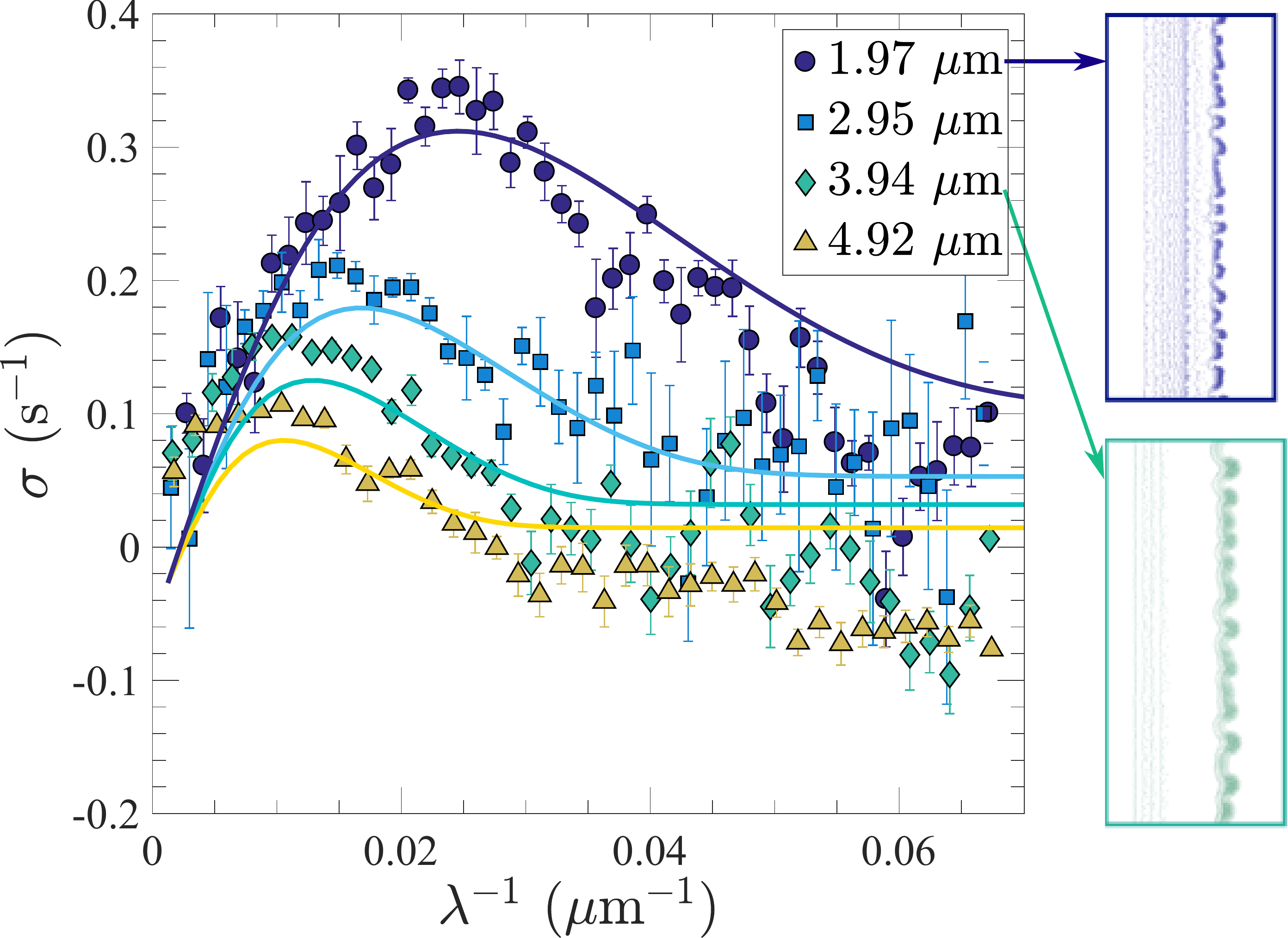}}
\hspace{0.5cm}
\subfloat[][]{\includegraphics[width=0.47\textwidth]{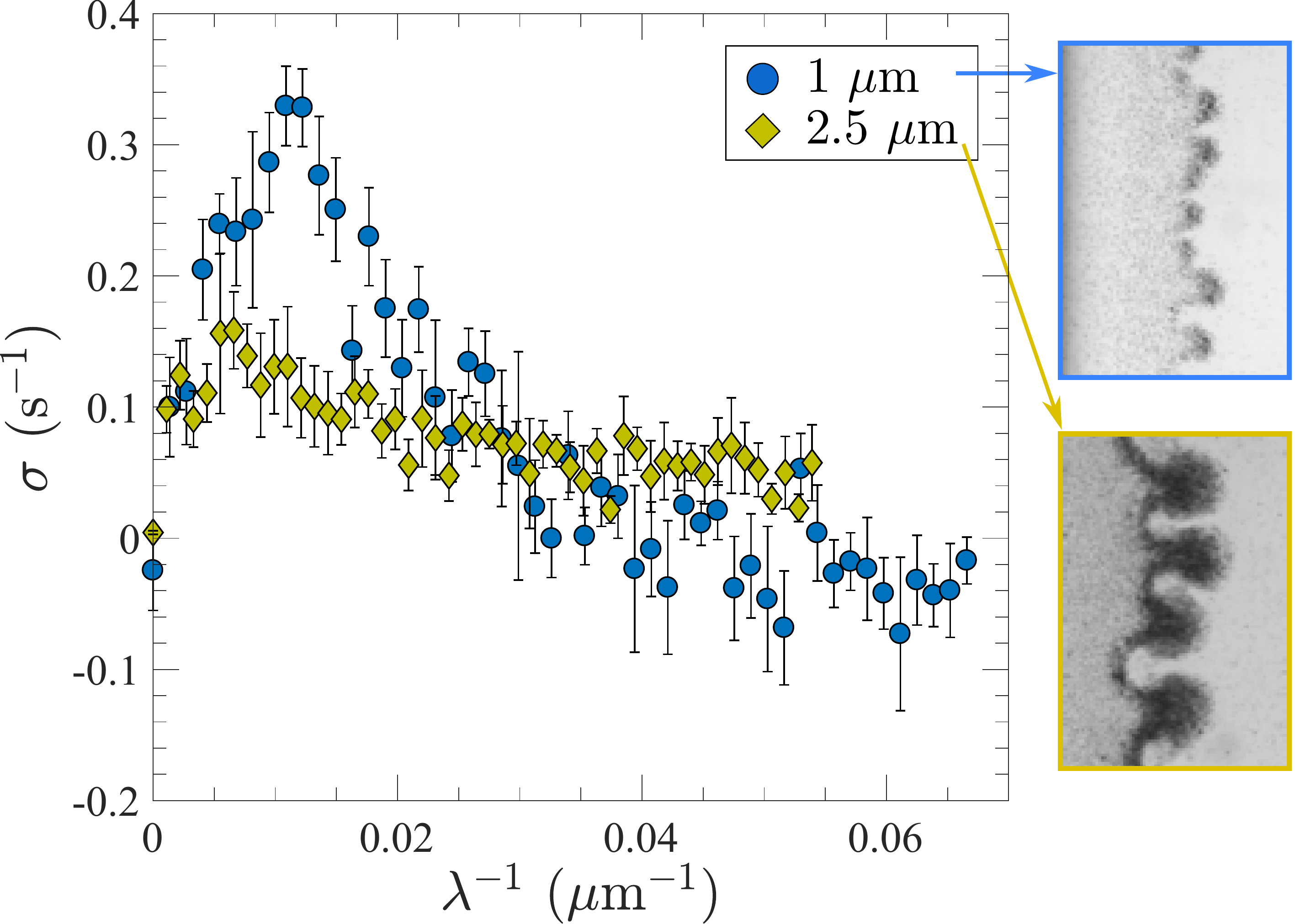}}
\caption{(a) Stability diagram. Symbols: quasi-2D simulations for $h=1.97-4.92$ $\mu$m. Solid lines:  theoretical predictions from the two-line model for the growth rate $\sigma_1$ for $d = 9.5h$. The pictures on the right show snapshots of the fingering instability at two different heights $h = 1.97, 3.94$ $\mu$m. (b) Experimental stability diagram for gravitational heights $h_g = a + k_BT/mg = 1, 2.5$ $\mu$m. The pictures on the right show snapshots of the fingering instability.} 
\label{fig:comparisons}
\end{figure}

Figure \ref{fig:comparisons}a compares the growth rate of the two line model with our quasi-2D particle simulations, where the microrollers are restricted to the $xy$-plane \cite{critters}, at several heights $h=1.97-4.92$ $\mu$m.  
Two parameters must be adjusted in the two-line model to match these simulations: the distance between the two lines $d$ to match $\lambda_m$, and the strength $S$ (or $\rho_0$) to match the magnitude of the maximum growth rate $\sigma_m$. Setting $d/h = 9.5$ matches $\lambda_m$ quantitatively for all the simulated heights, which shows that the instability wavelength depends linearly on $h$. It also suggest that, as shown by our previous work on the shock front \cite{Delmotte2017}, the width of the front is proportional to the height above the wall.  
The experimental stability diagrams in Fig.\ \ref{fig:comparisons}b are qualitatively similar: they have a well-defined fastest growing mode $\lambda_m$ which increases with the particle height.  

Thus the two-line model contains the essential physical ingredients to capture the fingering instability observed in the simulations and in the experiments. 
Its simplicity allows us to study the physical meaning of each term in the governing equations in order to understand the mechanisms at play.

\section{Mode selection by transverse flows}

\begin{figure}
\includegraphics[width=0.25\textwidth]{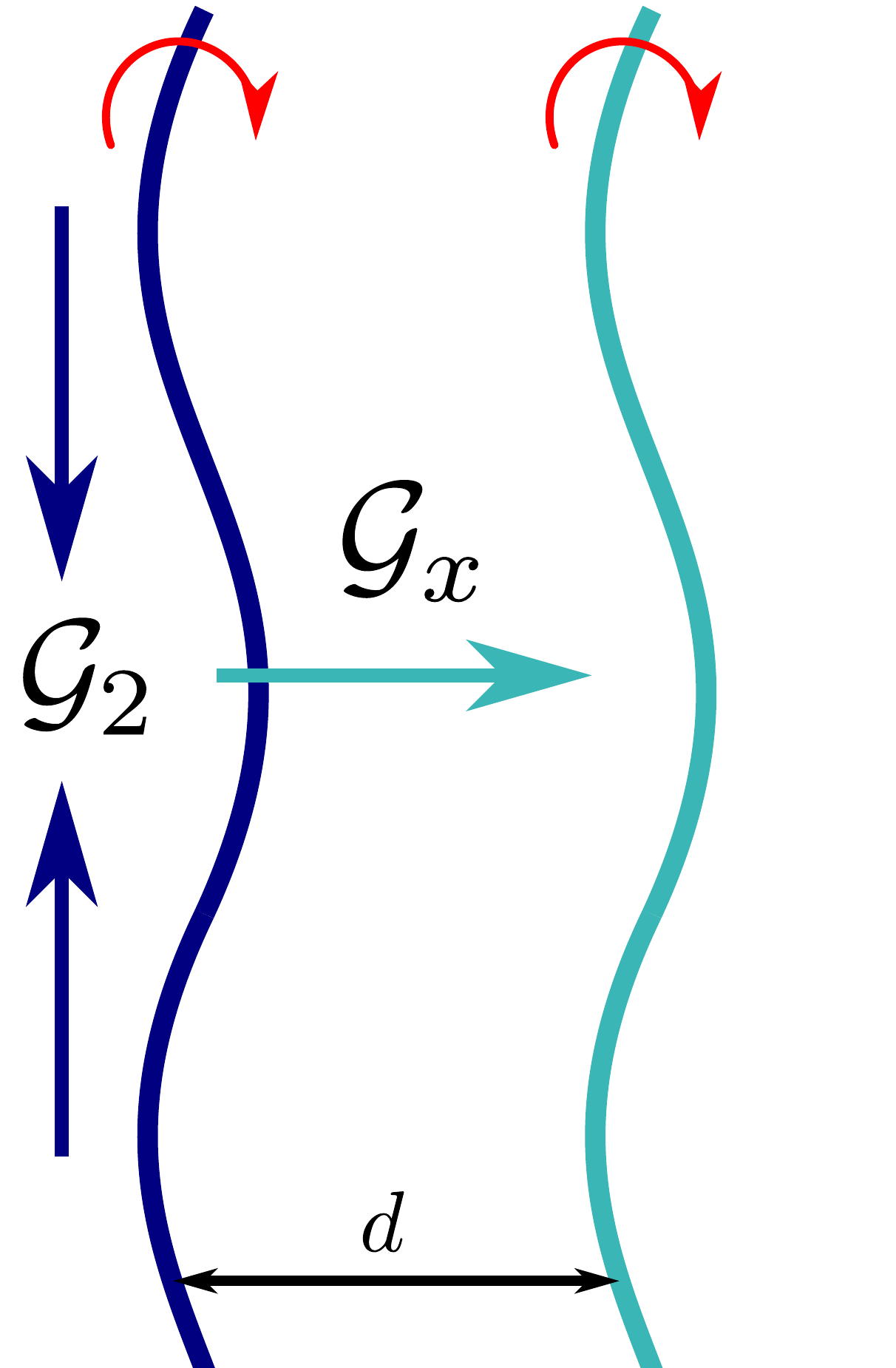}
\caption{Sketch of the main mechanisms of the fingering instability. The light blue arrows represent the  advection induced by the other line, corresponding to the term $\mathcal{G}_x$ in Eq. \eqref{eq:int_kernelx}. The dark blue arrows indicate the self-induced transverse motion, corresponding to the term $\mathcal{G}_2$ in Eq. \eqref{eq:G2}. The red  curved arrows represent the direction of rotation.}
\label{fig:sketch_mecanism}
\end{figure}

In this section we examine the role of each term from the matrix $A$ (Eq.\ \eqref{eq:MatA}) that appear in the full expression of the real positive eigenvalue $\sigma_1(k)$ in Eq.\ \eqref{eq:EVS}. We find that the location and the shape of the peak around $\lambda = \lambda_m$ in $\sigma_1$  are mainly controlled  by the following term in Eq.\ \eqref{eq:EVS}
\begin{eqnarray}
\left(A_{14}A_{31}\right)^{1/2}  &=& \left(-k\rho_0^2\tilde{\mathcal{G}}_2(k) \tilde{\mathcal{G}}_x(-d,k) \right)^{1/2} \\
&=& \frac{\sqrt{2}}{3}Sd\rho_0k^{5/2}\left(\frac{hK_1(2hk)K_2(\sqrt{d^2+4h^2}k)}{d^2+4h^2} \right)^{1/2},
\end{eqnarray}
where $K_n(x)$ is the modified Bessel function of the second kind.
Without this term there is no peak and therefore no lengthscale selection. The product $\tilde{\mathcal{G}}_2(k) \tilde{\mathcal{G}}_x(-d,k)$ indicates that both terms $\mathcal{G}_x$ and $\mathcal{G}_2$ must be nonzero to generate the fingering instability. 
$\mathcal{G}_x$ corresponds to the displacement of one line $\delta x_{i}$ induced by the density perturbations of the other line $\delta\rho_j$. $\mathcal{G}_2$ corresponds to the transverse velocity in the $y$ direction that drives the density perturbations $\delta\rho_i$ on each line due to the displacement on that line  $\delta x_i$. A sketch of these mechanisms is shown in Figure \ref{fig:sketch_mecanism}. The physical meaning of this product is that \textit{both} advection by the other line and self-induced transverse motion must be present to generate the fingering instability. 
We therefore conclude that the transverse instability is due to both the advective and transverse flows induced by a particle rotating parallel to a no-slip boundary (see Fig.\ \ref{fig:sketch_lines}b). 
Without both these advective and transverse flows, \modif{we would not observe this type of linear fingering instability}. \\

To demonstrate this we now consider a flat stress-free surface, i.e.\ a free-slip boundary such as an air-water interface with small curvature.  The image system of a rotlet near a flat slip surface is a counter rotating rotlet with the same magnitude \cite{DiLeonardo2011}, which ensures zero tangential stress at the interface. The corresponding velocity kernels replacing \eqref{eq:int_kernelx}-\eqref{eq:int_kernely} are
\begin{eqnarray}
 v_{x}\left(x,y;h\right)  &=& \mathcal{H}_x(x-x',y-y';h)\nonumber\\ 
 &=& -\dfrac{Sh}{3}\dfrac{1}{\left[(x-x')^2 + (y-y')^2 + 4h^2\right]^{3/2}}, \\
 v_{y}\left(x,y;h\right)  &=& \mathcal{H}_y(x-x',y-y';h) =  0.\label{eq:vy_slip}
\end{eqnarray}
The absence of transverse flow in \eqref{eq:vy_slip} leads to the absence of peak in the growth rate, \modif{i.e.\ no fastest growing mode,} and thus no clear lengthscale selection for the two-line model, see Appendix B for a detailed analysis. Therefore, microrollers near a slip surface should not be subject \modif{to a linear fingering instability with a well-defined finger width}.



\section{Conclusions and discussion}
We derived the simplest possible model that can capture the fingering instability observed both experimentally and numerically in suspensions of microrollers above a floor. Our model directly accounts for the nonlocal hydrodynamic interactions between the particles.  Our analytic linear stability analysis confirmed that the fingering instability is linear, that it happens in the plane parallel to the floor\Michelle{,} and is purely hydrodynamic in origin \footnote{We also carried 3D simulations of rotating particles with no external force above a no-slip surface and still observed the fingering instability, thus confirming the hydrodynamic origin of the instability.}. Our comparisons with quasi-2D particle simulations showed quantitative agreement for a range of heights by adjusting only the distance  between the two lines as a function of the height to  $d=9.5h$. These results showed that the instability wavelength and the front width depend linearly on $h$. 
Thanks to the simplicity of the model, we identified each term separately and showed that the transverse flows due to the nearby no-slip surface are responsible for the lengthscale selection. In particular, for a free-slip bottom wall, there are no transverse flows and there is no linear instability. 
The two-line model and the linear stability analysis have improved our understanding of the fingering instability and shed more light on the genesis of the autonomous motile colloidal structures called ``critters".

Our two-line model considers only far-field hydrodynamic interactions and does not account for finite particle size effects. One can test the effect of near-field hydrodynamics on the qualitative behaviour of the system with numerical simulations. In the case of the microroller fingering instability, we found that far-field hydrodynamics alone achieved  qualitative agreement with the experiments.


More importantly, our two-line model is not limited to rotating particles and finds interesting applications in other particulate systems. It could  be used to study the sedimentation of particles parallel to a wall or to a slip surface, but also to investigate the formation of phonons in microfluidic systems of confined driven particles (e.g. droplets \cite{Beatus2006} or microswimmers \cite{Tsang2016b}). Indeed, both kind of particles generate transverse flows that   destabilize the systems and could be analyzed within the same framework. One would only have to change the velocity kernels (and regularize them when necessary): stokeslets and their image system for sedimenting particles \cite{Blake1974},  source dipoles for confined driven particles \cite{Beatus2012, Brotto2013}. Surprisingly, the sedimentation of particles adjacent to a single wall has received little attention in the literature. Our preliminary large scale particle simulations suggest that the parallel wall selects a characteristic wavelength in the unstable sedimenting suspensions. We will study this phenomenon experimentally, numerically and theoretically in more detail in a near future.

\section{Acknowledgments}
This work was supported primarily by the Materials Research Science and Engineering Center (MRSEC) program of the National Science Foundation under Award Number DMR-1420073 and the Gordon and Betty Moore Foundation through Grant GBMF3849. A. Donev and B. Delmotte were supported in part by the National Science Foundation under award DMS-1418706. P. Chaikin was partially supported by the Center for Bio-Inspired Energy Science, A DOE BES EFRC under Award DE-SC0000989. We gratefully acknowledge the support of the NVIDIA Corporation with the donation of GPU hardware for performing some of the simulations reported here.

\appendix
\section{Entries and eigenvalues of the matrix A}

\noindent The matrix $A$ contains only six distinct coefficients listed below:
\begin{flalign}
 A_{11} &= -A_{22} && \nonumber\\
        &= \rho_0\mathcal{G}_0 &&\nonumber\\
        &=   \rho_0 Sh \dfrac{8d(d^2-4h^2)}{3[d^2 + 4h^2]^{3}} &&
\end{flalign}
\begin{flalign}
 A_{12} &= -A_{21} &&\nonumber\\
        &= \rho_0\tilde{\mathcal{G}}_1(k) &&\nonumber\\
        &=  \rho_0\int\limits_{-\infty}^{\infty} \mathcal{G}_1(y)\exp(-iky)dy && \nonumber\\
        &= \rho_0Sh\dfrac{2d}{15}\left[\dfrac{-3d^2+8h^2}{(d^2+4h^2)^{3/2}}k^3 K_3\left(\sqrt{d^2+4h^2}k\right)\right. &&\nonumber\\
      &\left.+ \dfrac{8}{(d^2+4h^2)^2}MG\left(\left\lbrace(-1/2),()\right\rbrace,\left\lbrace(0,2),(1/2)\right\rbrace,\dfrac{1}{4}(d^2+4h^2)k^2\right) \right] &&
\end{flalign}
where $K_n(x)$ is the modified Bessel function of the second kind, and $MG$ is the Meijer G-function.
\begin{flalign}
 A_{14} &= A_{23} &&\nonumber\\
        &= \tilde{\mathcal{G}}_x(-d,k) &&\nonumber\\
        &=  Sh\dfrac{2d^2}{3(d^2+4h^2)}k^2 K_2\left(\sqrt{d^2+4h^2}k\right) &&
\end{flalign}
\begin{flalign}
 A_{31} &= A_{42} &&\nonumber\\
        &= -ik\rho_0^2\tilde{\mathcal{G}}_2(k) &&\nonumber\\
        &= \rho_0^2Sh\dfrac{2}{6h}k^3 K_1\left(2hk\right) &&
\end{flalign}
\begin{flalign}
 A_{32} &= A_{41} &&\nonumber\\
        &= -ik\rho_0^2\tilde{\mathcal{G}}_3(k) &&\nonumber\\
        &= \rho_0^2Sh\dfrac{2}{15}\left[\dfrac{4d^2-4h^2}{(d^2+4h^2)^{3/2}}k^4 K_2\left(\sqrt{d^2+4h^2}k\right)\right. &&\nonumber\\
      &\left.- \dfrac{4}{(d^2+4h^2)^2}k MG\left(\left\lbrace(-1),()\right\rbrace,\left\lbrace(1/2,3/2),(0)\right\rbrace,\dfrac{1}{4}(d^2+4h^2)k^2\right) \right] &&
\end{flalign}
\begin{flalign}
 A_{34} &= -A_{43} &&\nonumber\\
        &= -ik\rho_0\tilde{\mathcal{G}}_y(-d,k) &&\nonumber\\
        &= \rho_0Sh\dfrac{2d}{3\sqrt{d^2+4h^2}}k^3 K_1\left(\sqrt{d^2+4h^2}k\right) &&
\end{flalign}
The structure of the matrix $A$ permits an analytic calculation of the eigenvalues:
\begin{eqnarray}
\sigma_{1,2,3,4}(k) &=& \pm\frac{1}{2}\left[2A_{11}^2 - 2A_{12}^2 + 4A_{32}A_{14}-2A_{34}^2\right. \nonumber\\
& &\pm 2\left(A_{11}^4-2A_{11}^2A_{12}^2+4A_{11}^2A_{14}A_{32}+2A_{11}^2A_{34}^2-8A_{11}A_{14}A_{31}A_{34} +A_{12}^4\right. \nonumber\\
&& \left. \left.-4A_{12}^2A_{14}A_{32}-2A_{12}^2A_{34}^2+8A_{12}A_{14}A_{32}A_{34}+4A_{14}^2A_{31}^2-4A_{14}A_{32}A_{34}^2+A_{34}^4 \right)^{1/2} \right]^{1/2}.\nonumber\\
\label{eq:EVS}
\end{eqnarray}

\section{Microrollers above a free-slip surface}

\begin{figure}
\subfloat[][]{\includegraphics[width=0.35\textwidth]{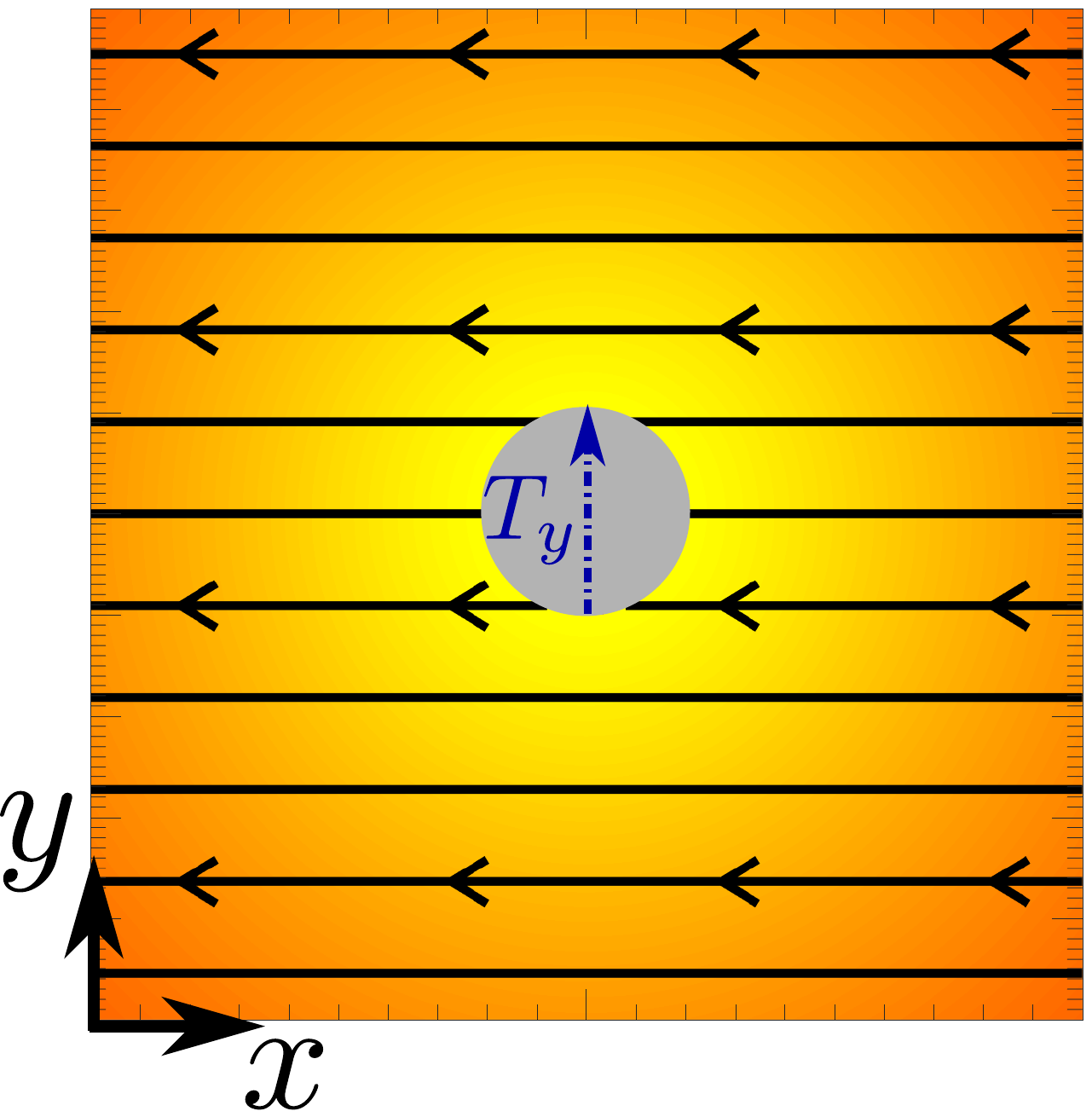}}
\hspace{0.5cm}
\subfloat[][]{\includegraphics[width=0.55\textwidth]{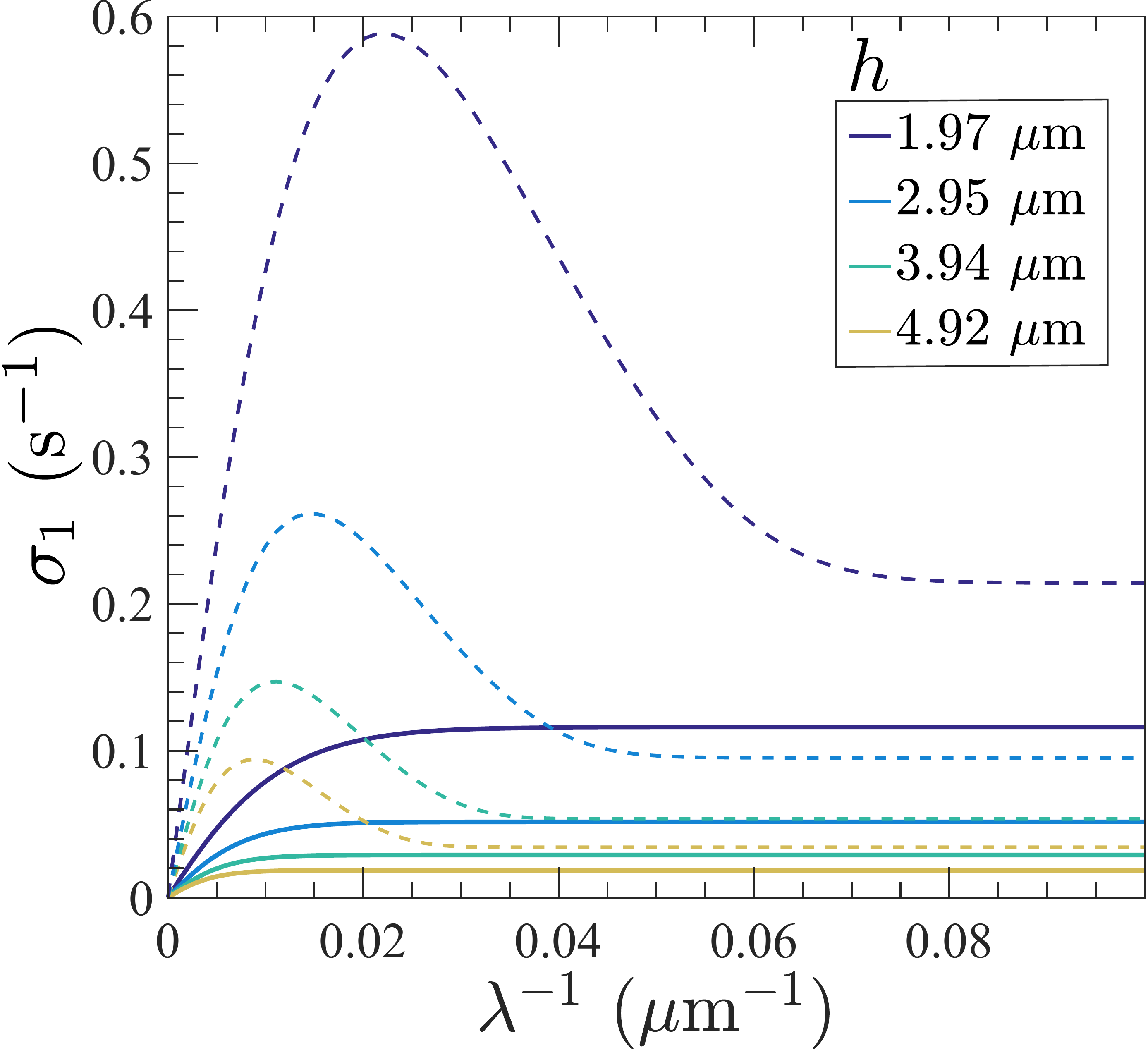}}
\caption{(a) Flow field around a rotlet in the plane of rotation above a slip surface. The arrow indicates the direction of the applied torque $T_y$. The colors, from red to yellow, represent the magnitude of the in-plane velocity. (b) Growth rate $\sigma_1$ vs. $\lambda^{-1}$ for the two line model at various height $h = 1.97-4.92$ $\mu$m and $d/h=10$. Dashed line: above a no-slip surface. Solid line: above a slip surface.} 
\label{fig:slip_vs_no_slip}
\end{figure}

Figure \ref{fig:slip_vs_no_slip}a shows the flow field around a particle above a  slip  surface. Note that the streamlines in the plane of rotation are parallel, and that the induced translational velocity is in the opposite direction to the no-slip case (see Fig.\ \ref{fig:sketch_lines}b). 
After carrying the same linear stability analysis we obtain the following matrix:
$$ A_{\mbox{slip}}=  \left(\begin{array}{cccc}
\rho_0\mathcal{H}_0 & \rho_{0}\tilde{\mathcal{H}}_1 & 0 & \tilde{\mathcal{H}}_{x}(-d,k)\\
-\rho_{0}\tilde{\mathcal{H}}_1 & -\rho_0\mathcal{H}_0 & \tilde{\mathcal{H}}_{x}(-d,k) & 0\\
0 & 0 & 0 & 0\\
0 & 0 & 0 & 0
\end{array}\right), \label{eq:MatB}$$
where 
$$ \mathcal{H}_0 = -\frac{4}{3}Sh\frac{d}{(d^2+4h^2)^2},$$
$$ \tilde{\mathcal{H}}_1 = \frac{2}{3}Sh\frac{d}{(d^2+4h^2)}k^2K_2(\sqrt{d^2+4h^2}k),$$
and 
$$ \tilde{\mathcal{H}}_x = -\frac{2}{3}Sh\frac{k}{(d^2+4h^2)}K_1(\sqrt{d^2+4h^2}k).$$
The nonzero eigenvalues of the new matrix $A_{\mbox{slip}}$ are
\begin{eqnarray}
 \sigma_{1,2}(k) &=& \pm\rho_0\left[\mathcal{H}_0^2-\tilde{\mathcal{H}}_1^2\right]^{1/2},
\end{eqnarray}
and 
\begin{eqnarray}
 \lim_{k\rightarrow\infty} \sigma_{1,2}(k) &=& \pm\rho_0\mathcal{H}_0,
\end{eqnarray}
Figure \ref{fig:slip_vs_no_slip}b compares the stability diagram between a slip and a no-slip surface for $h=1.97-4.92$ $\mu$m. The peak  around $\lambda_m$ disappears in the absence of  transverse flows, meaning that there is no clear lengthscale selection above a slip surface. The constant term $\rho_0\mathcal{H}_0$ in the matrix $A_{\mbox{slip}}$ plays the same role as $\rho_0\mathcal{G}_0$, as shown by the plateaus  at short wavelengths.

\newpage

\bibliographystyle{unsrt}

\bibliography{Paper_theory_instability}

\end{document}